\documentstyle[preprint,aps,eqsecnum]{revtex}
\newcommand{\beq}[1]{\begin{equation}\label{#1}}
\newcommand{\eeq}{\end{equation}}
\newcommand{\bear}[1]{\begin{eqnarray}\label{#1}}
\newcommand{\ear}{\end{eqnarray}}
\newcommand{\nn}{\nonumber}

\newcommand{\rf}[1]{(\ref{#1})}
\newcommand{\nl}{ {\hfill \break} }

\renewcommand{\ol}{ \overline}

\newcommand{\R}{ \mbox{\rm I$\!$R} }
\def\C{\mbox{\rm {I\kern-.520em C}}}
\newcommand{\End}{ {\rm End} }

\newcommand{\E}{ {\rm E} }
\newcommand{\BD}{ {\rm BD} }

\newcommand{\pr}{ {\rm pr} }

\newcommand{\partlx}[1]{\frac{\partial}{\partial x^{#1}}}

\newcommand{\p}{\partial}

\newcommand{\sq}[1]{\sqrt{|#1|}}

\newcommand\mustbe{\stackrel{!}{=}}

\begin{document}
\draft
%
\title{Einstein and Brans-Dicke frames
in multidimensional cosmology}
\author{M. Rainer 
\thanks{E-mail: mrainer@phys.psu.edu}
} 
\address{
Gravitationsprojekt, Mathematische Physik I,
Institut f\"ur Mathematik,Universit\"at Potsdam, 
\\
PF 601553, D-14415 Potsdam, Germany
\\
and
Center for Gravitational Physics and Geometry,
104 Davey Laboratory, The Pennsylvania State University,
University Park, PA 16802-6300, USA
}
\author{A. Zhuk
\thanks{E-mail: zhuk@paco.odessa.ua}
}
\address{
Gravitationsprojekt, Mathematische Physik I,
Institut f\"ur Mathematik,Universit\"at Potsdam, 
\\
PF 601553, D-14415 Potsdam, Germany
\\
and
Department of Physics,
University of Odessa, 2 Petra Velikogo St.,
Odessa 270100, Ukraine
}
\maketitle
\vspace*{-0.7cm}
\begin{abstract}
Inhomogeneous multidimensional cosmological models
with a higher dimensional  space-time manifold
$M={\ol M}_0\times\prod_{i=1}^n M_i$ ($n\geq 1$) are investigated
under dimensional reduction to a ${D}_0$-dimensional
effective non-minimally coupled $\sigma$-model
which generalizes the familiar Brans-Dicke
model.
It is argued that the Einstein frame should be considered
as the physical one.
The general prescription for the Einstein frame reformulation
of known solutions in the Brans-Dicke frame is given.
As an example, the reformulation is demonstrated explicitly
for the generalized Kasner solutions where
it is shown that in the Einstein frame there are no
solutions with inflation of the external space.
\end{abstract}
%
\section{\bf Introduction}
\setcounter{equation}{0}
All contemporary  unified interaction models face
the requirement also to incorporate
gravity.
The most prominent attempt in this direction is
string theory and its recent extension of M-theory
\cite{SV,Du}
which extends strings to generalized membranes
as higher-dimensional  objects.
Most of these unified models
are modeled initially on a higher-dimensional space-time manifold,
say of dimension $D>4$, which then
undergoes some scheme of spontaneous compactification
yielding a direct product manifold $M^4\times K^{D-4}$
where $M^4$ is the manifold of space-time
and $K^{D-4}$ is a compact internal space
(see e.g. \cite{SchSch}-\cite{DNP}).
Hence it is natural to investigate cosmological consequences
of such a hypothesis.

In particular we will investigate multidimensional cosmological
models (MCM) given as a topological product
\beq{1.1}
M=\overline{M}_{0}\times\prod_{i=1}^{n} M_{i} ,
\eeq
where $\overline{M}_{0}:=\R\times M_0$
is a ${D}_0$-dimensional (usually ${D}_0=4$)
smooth space-time manifold with spatial sections
all diffeomorphic to a standard section $M_0$,
and $\prod_{i=1}^{n} M_{i}$ an internal
product space from smooth homogeneous factor spaces $M_i$
of dimension $d_i$, $i=1,\ldots,n$.

Let $\overline{M}_{0}$ be equipped with a smooth
hyperbolic metric $\overline{g}^{(0)}$,
let $\ol\gamma$ and $\beta^i$ , $i=1,\ldots,n$
be smooth scalar fields on $\overline{M}_{0}$, and
let each $M_i$ be equipped with a smooth homogeneous metric $g^{(i)}$.
Then,
under any projection $\pr: M\to \overline{M}_{0}$
a pullback
consistent with \rf{1.1}
of $e^{2\ol\gamma}\overline{g}^{(0)}$
from $x\in \overline{M_{0}}$ to $z\in\pr^{-1}\{x\}\subset M$
is given by
\beq{1.2}
g(z):=e^{2\ol\gamma(x)}\overline{g}^{(0)}(x)
+\sum_{i=1}^{n} e^{2\beta^i(x)}g^{(i)} .
\eeq
The function $\ol\gamma$
fixes a {\em gauge} for the (Weyl)
{\em conformal frame} on $\overline{M_0}$.
Note that the latter has little in common with a usual
(coordinate) frame of reference.
Rather it corresponds to a particular choice
of geometrical variables,
whence it might also be called a (classical) {\em representation}
of the metric geometry. All these terms are often used synonymously
in the literature, and so we do below.

We will show below how $\ol\gamma$ uniquely defines
the form of the effective ${D}_0$-dimensional theory.
For example $\ol\gamma:=0$ defines the
Brans-Dicke frame\footnote{This frame is sometimes also called
Brans-Dicke-Jordan frame, or simply Jordan frame.}
with a non-minimally coupled dilatonic
\footnote{Here by a {\em dilatonic} scalar field
we refer to any scalar field which is given in
terms of logarithms of internal space scale factors.}
scalar field
given
by the total internal space volume,
while (for$ {D}_0\neq 2$)
$\ol\gamma:=\frac{1}{2-{D}_0}\sum_{i=1}^{n}d_i\beta^i$
defines the Einstein frame
\footnote{This frame is sometimes also called
Einstein-Pauli frame, or simply Pauli frame.}
with all dilatonic  scalar fields minimally coupled.

There is a long and still ongoing
( see e.g. \cite{Di} )
discussion in the literature
which frame is the physical one.
Historical references on this subject are contained
in \cite{MaSo}, and more recent ones in \cite{RZ}.

{}From the mathematical point of view,
the equivalence of all classical representations
of smooth geometrical models based on
multidimensional metrics \rf{1.2} related
by different choices of the smooth gauge function $\ol\gamma$
is guaranteed by the manifest regularity of the conformal
factor $e^{2\ol\gamma}>0$.
Hence  the spaces of  regular and smooth
local classical solutions are isomorphic
for all regular and smooth representations
of the classical geometrical theory.
Note however that physically interesting choices
of $\ol\gamma$ might sometimes fail to exist within
any class of functions which satisfies the
required regularity and smoothness
conditions.
So e.g. for ${D}_0=2$ a gauge of $\ol\gamma$
yielding the Einstein frame fails to exist,
whence some $2$-dimensional scalar-tensor theory obtained
by dimensional reduction from a multidimensional
geometry is in general not conformally equivalent
to a theory with minimal coupling.

Even if two classical conformal representations
are equivalent from the purely geometrical point,
their different coupling of a dilatonic scalar field
to the metric geometry in different conformal frames
distinguishes the representations physically,
if and only if
physics depends indeed on the metric geometry
rather than on the Weyl geometry only.

Moreover, if the theory incorporates additional matter fields,
the dynamics of these fields may reveal the true physical
frame to which they couple.

Let us point out in advance the main advantages
of the Einstein frame, corresponding to $f=0$ in \rf{2.5} below,
for the multidimensional model \rf{1.1}
with metric \rf{1.2}.

First, in this frame all dilatonic
fields have the same
(positive) sign in all kinetic terms.
In  other frames with $f\neq 0$ there is a dilatonic kinetic
term, $(\p f)^2$ in \rf{2.15}, which may have an opposite (negative) sign
corresponding to a ghost.
Hence there is no way to guarantee ``unitarity44 (i.e. the positive
definiteness of the Hamiltonian) for the action  \rf{2.15}
if one tries to identify a Brans-Dicke frame (with $f\neq 0$)
as a physical one \cite{Cho}.
Although the gauge $f=-2\ol\gamma$ also provides the correct
sign for all kinetic terms,
it has another drawback in its
coupling to additional scalar matter fields (see below) .

Secondly, Cho \cite{Cho} has also shown
that only in the Einstein frame the perturbative part
of the gravitational
interaction is generated purely by spin-$2$ gravitons.
In any Brans-Dicke frame additional spin-$0$ scalar particles
enter as basic perturbative modes of gravity.

Third, only in the Einstein frame the $D_0$-dimensional
effective gravitational constant
(which is Newton's constant for $D_0=4$)
is an exact constant, such that the present day
experimental bounds on the variation of the
gravitational constant \cite{Mar,KPW} are solved automatically,
while in all Brans-Dicke frames fine-tuning is necessary.

Related arguments in favor of the Einstein frame
given in \cite{MaSo} for $4$-dimensional non-linear
(higher order) gravitational models
may be applied analogously to scalar-tensor gravity theories
as the ones considered here.
There it was shown that
``the existence of the Einstein frame is in any case
essential for assessing classical stability of Minkowski space
and positivity of energy for nearby solutions.
In the Jordan frame, the dominant energy condition never holds.
For these reasons, the Einstein frame is the most natural candidate
for the role of physical frame''.
Note that while \cite{MaSo} discussed
a generic possibility
for (multi-)scalar-tensor theories
to couple extra scalar fields at hand to the dilatonic one
and to the metric in any conformal frame
for our $D_0$-dimensional effective theory
this possibility  does not arise.\footnote{
In this aspect our approach differs also essentially
from that of \cite{GGB}, who do not consider scalar fields
and their couplings as given by reduction from a higher-dimensional
space, whence from that point of view it is still consistent when
they favor a Brans-Dicke frame.}
Here a choice of gauge function $\ol \gamma$
not only fixes a conformal frame and the dilatonic field $f$
but also all the couplings with further dilatonic scalar fields, 
prescribed then by the
particular multidimensional structure of
our $D$-dimensional theory.

Below we take this higher-dimensional theory
in form of an Einsteinian theory plus any
minimally coupled $D$-dimensional
matter which is in accordance with ansatz given by \rf{1.1}
and \rf{1.2} homogeneous in the $(D-D_0)$ - dimensional internal
space, i.e. all free functions only depend on $\overline{M_0}$
(like e.g. the zero mode fields in \cite{GSZ}).

With the above multidimensional structure,
this ansatz fixes also all couplings of this extra matter
to geometrical fields in the effective $D_0$-dimensional theory.
It is evident from \rf{2.15} below that the extra matter
is also minimally coupled to ${\overline g}^{(0)}$ for any gauge of
$\ol\gamma$,
but its coupling to the dilatonic field  strongly varies with $\ol\gamma$.

In a Brans-Dicke frame (where $\ol \gamma=0$)
matter couples directly to a dilatonic prefactor
$e^f=\prod_{i=1}^{n} a_i^{d_i}$
(in front of kinetic as well as potential terms )
which is
proportional to the Riemann-Lebesgue volume of the total internal space.

In the Einstein frame ($f=0$) dilatonic fields become,
like the extra matter,
minimally coupled to the geometry of ${\overline M}_{0}$.
Then, the extra matter is coupled to dilatonic fields
by via a potential term of the effective $D_0$-dimensional theory.

So, in the Einstein frame the physical setting is rather clear:

First, with respect to scalar fields of dilatonic origin
the theory has the shape of a self-gravitating $\sigma$-model
\cite{RZ}
with flat Euclidean target space
and self-interaction described by an effective potential.
Eventually existing minima of this potential
have been identified as positions admitting a stable 
compactification.\footnote{The stability analysis of the compactified 
internal spaces in multidimensional cosmological models \cite{Ma} as well 
as multidimensional black hole solutions \cite{BM} has also been performed 
in the Einstein frame.} 
Small fluctuations of scale factors of the internal spaces near such minima 
could in principle be observed as massive scalar fields (gravitational 
excitons) in the external space-time
\cite{GZ}.

Second, under the assumption that the fluctuations
of the internal space scale factors
around a stable position
at one of the minima mentioned above
are very small,
the extra matter fields (of any type) might
be considered in an approximation of order zero
in these fluctuations. In this approximation they
have the usual free $D_0$-dimensional form
and follow the geodesics of the metric
${g}^{(\E)}:={\overline g}^{(0)}\vert_{f=0}$.
Taking into account the first nontrivial order in these
fluctuations yields the gravitational excitons plus an
interaction between the extra matter and the excitons
\cite{GSZ}. 
A likewise clear structure is not at hand for the corresponding theory 
in a Brans-Dicke frame.

Besides these arguments in favor of the Einstein frame,
it interesting to note that many  investigations
of astrophysical consequences for scalar-tensor theories
are also performed in the Einstein frame as the physical one
\cite{CSY,MR,SSM}.

For cosmological models with
multidimensional structure \rf{1.1} and metric structure \rf{1.2}
most exact solutions of the field equations were obtained
in the spatially homogeneous case, where the scale factors
$a_i:=e^{\beta^i}$, $i=1,\ldots,n$ are only a function of
time $t\in\R$.
Some overview and an extensive list of references is given in
\cite{IM,GIM,KZ,KRZ}.
All solutions known to us have been obtained exploiting the
simple coupling ($\ol\gamma=0$) in the
Brans-Dicke frame. However the arguments above
show that these solutions should be reformulated
in the Einstein frame before a physical interpretation
is given. It is clearly to be expected that the reinterpreted solutions
will have a different qualitative behavior
as compared to those in the Brans-Dicke frame.
The concretization of this expectation is our major motivation
for the present investigations.
The common underlying structure of many exact solutions
rises the possibility to find an explicit
description of the transition from Brans-Dicke to Einstein frame
for rather general classes of solutions.

As an important example,
the exact transformation can be performed for
the well known generalized Kasner solution.
The so obtained solution in the Einstein frame is
indeed qualitatively quite different than the Kasner one,
which conclusively supports our previous expectations.

The paper is organized as follows:
In Sec. \ref{Sect. 2} we describe the multidimensional  model
and obtain a dimensionally reduced effective theory
in an arbitrary frame.
Sec. \ref{Sect. 3} presents a general method for transformations between solutions
in Brans-Dicke and Einstein frames.
A brief review of the generalized Kasner solution in the
Brans-Dicke frame is given in Sec. \ref{Sect. 4}.
Its explicit reformulation in the Einstein frame
follows in Sec. \ref{Sect. 5}.

\section{\bf Multidimensional geometry as effective $\sigma$-model}
\label{Sect. 2}
\setcounter{equation}{0}
Let us now consider a multidimensional manifold
\rf{1.1} of dimension $D={D}_0 + \sum_{i=1}^{n} d_i
= 1 + \sum_{i=0}^{n} d_i $,
equipped with
a (pseudo) Riemannian metric
\rf{1.2} where
\beq{2.1}
g^{(i)} \equiv g_{m_{i} n_{i}}(y_i) dy_i^{m_{i}} \otimes dy_i^{n_{i}} ,
\eeq
are  $R$-homogeneous Riemannian metrics on $M_i$
(i.e. the Ricci scalar $R[g^{(i)}] \equiv R_i$ is a constant on $M_i$),
in coordinates $y_i^{n_{i}}$,
$n_{i}=1,\ldots, d_{i}$,
and
\beq{2.2}
x\mapsto \ol{g}^{(0)}(x)
=\ol{g}^{(0)}_{\mu\nu}(x) d x^{\mu} \otimes d x^{\nu}
\eeq
yields a general, not necessarily $R$-homogeneous,
(pseudo) Riemannian metric on $\ol{M}_0$.

Below,
the $\overline{g}^{(0)}$-covariant derivative of a given function $\alpha$
w.r.t. $x^\mu$ is denoted  by  $\alpha_{;\mu}$,
its partial derivative also by $\alpha_{,\mu}$,
and $(\p\alpha)(\p\beta):=\ol{g}^{(0)\mu\nu} \alpha_{,\mu} \beta_{,\nu} $.
Furthermore we use the shorthand
$|g| := |\det (g_{MN})|$,
$|\ol{g}^{(0)}| := |\det (\ol{g}^{(0)}_{\mu\nu})|$,
and analogously for all other metrics including $g^{(i)}$, $i=1, \ldots, n$.

On $\overline{M}_0$, the Laplace-Beltrami operator
$\Delta[\overline{g}^{(0)}]=
\frac{1}{\sqrt{|\overline{g}^{(0)}|}}
{\partlx{\mu}}
\left( \sqrt{|\overline{g}^{(0)}|} \overline{g}^{(0)\mu\nu}
{\partlx{\nu}}
\right)
$ ,
transforms under the conformal map
$\overline{g}^{(0)}\mapsto e^{2\overline{\gamma}} \overline{g}^{(0)}$
according to
\bear{2.3}
\Delta[e^{2\overline{\gamma}}\overline{g}^{(0)}]&=&
e^{-2\overline{\gamma}}\Delta[\overline{g}^{(0)}]
-e^{-2\overline{\gamma}}
{\overline{g}^{(0)}}^{\mu\nu}
\left(
\Gamma[e^{2\overline{\gamma}}\overline{g}^{(0)}]-\Gamma[\overline{g}^{(0)}]
\right)^{\lambda}_{\mu\nu}
{\partlx{\lambda}}
\nn\\
&=&
e^{-2\ol{\gamma}}
\left(
\Delta[\ol{g}^{(0)}]
+({D}_0-2){g^{(0)}}^{\mu\nu}
\frac{\partial \ol\gamma}{\partial x^\mu}
{\partlx{\nu}}
\right) ,
\ear
where $\Gamma$ denotes the Levi-Civita connection.

Then, for the multidimensional metric (\ref{1.2})
the Ricci tensor decomposes likewise into blocks
and the corresponding Ricci curvature scalar
reads
\bear{2.4}
R[g] &=&
e^{-2\ol{\gamma}}R[\ol{g}^{(0)}]
+\sum_{i=1}^{n} e^{-2\beta^i} R[g^{(i)}]
- e^{-2\ol{\gamma} }
\ol{g}^{(0)\mu\nu}
\left(
({D}_0-2)({D}_0-1)
\frac{\partial\ol{\gamma}}{\partial x^\mu}
\frac{\partial\ol{\gamma}}{\partial x^\nu}
\right.
\nn\\
& &
\left.
+ \sum_{i,j=1}^{n} (d_i\delta_{ij}+d_i d_j)
\frac{\partial\beta^i}{\partial x^\mu}
\frac{\partial\beta^j}{\partial x^\nu}
+2({D}_0-2)\sum_{i=1}^{n}d_i
\frac{\partial\ol{\gamma} }{\partial x^\mu}
\frac{\partial\beta^i}{\partial x^\nu}
\right)
\nn\\
& &
- 2 e^{-2\ol{\gamma} }
\Delta[\ol{g}^{(0)}]
\left(
({D}_0-1)\ol{\gamma}
+\sum_{i=1}^{n} d_i\beta^i
\right) .
\ear
Let us now set
\beq{2.5}
f \equiv {f}[\ol{\gamma}, \beta]
:= ({D}_0 - 2) \ol{\gamma} +\sum_{j=1}^{n} d_j  \beta^j ,
\eeq
where $\beta$
is the vector field with the dilatonic scalar fields
$\beta^i$ as components.
(Note that $f$ can be resolved for
$\ol\gamma\equiv \ol\gamma[f,\beta]$
if and only if ${D}_0 \neq 2$.
The singular case $D_0=2$ is discussed in \cite{RZ}.)
Then,
\rf{2.4} can also be written as
\bear{2.6}
 R[g] \! &\!-\! &\!
e^{-2\ol{\gamma} } R[\ol{g}^{(0)}]
-\sum_{i=1}^{n} e^{-2\beta^i} R_i \ =
\\\nn
= \!&\!-\! &\!  e^{-2\ol{\gamma} }
\left\{
\sum_{i=1}^{n} d_i (\p\beta^i)^2
+ (\p f)^2
+(D_0-2) (\p\ol{\gamma})^2
+ 2 \Delta[\ol{g}^{(0)}] (f+\ol{\gamma})
\right\}
\\\nn
=\! &\!-\! &\!  e^{-2\ol{\gamma}}
\left\{
\sum_{i=1}^{n} d_i (\p\beta^i)^2
+ ({D}_0 - 2) (\p \ol{\gamma})^2 - (\p f) \p (f + 2 \ol{\gamma})
+ R_{B} \right\} ,
\\
\label{2.7}
R_B &:=& \frac{1}{\sq {\ol{g}^{(0)}} } e^{-f}
     \p_{\mu} \left[2 e^f \sq {\ol{g}^{(0)} }
     {\ol{g}^{(0)\mu \nu}} \p_{\nu} (f + \ol{\gamma}) \right] ,
\ear
where the last term will yield just a boundary contribution \rf{2.12} 
to the action \rf{2.11} below.

Let us assume all $M_{i}$, $i=1,\ldots,n$, to be connected and oriented.  
Then a volume form on $M_i$ is defined by
\beq{2.8}
\tau_i  := \sqrt{|g^{(i)}(y_i)|}
\ dy_i^{1} \wedge \ldots \wedge dy_i^{d_i} ,
\eeq
and the total internal space volume is
\beq{2.9}
\mu:= \prod_{i=1}^n \mu_i, \quad
\mu_i := \int_{M_i} \tau_i =\int_{M_i} d^{d_i}y_i \sq{g^{(i)} } .
\eeq
If all of  the spaces $M_i$, $i=1,\ldots,n$ are compact,
then the volumes $\mu_i$ and $\mu$ are finite,
and so are also the numbers
$\rho_i=\int_{M_i}d^{d_i}y \sq{g^{(i)}} R[g^{(i)}] $.
However, a non-compact $M_i$ might have infinite volume $\mu_i$  or infinite
$\rho_i$.
Nevertheless, by the $R$-homogeneity of $g^{(i)}$
(in particular satisfied for Einstein spaces),
the ratios
$\frac{\rho_i}{\mu_i}=R[g^{(i)}] $,
$i=1,\ldots,n$, are just finite constants.
In any case, we must tune the
$D$-dimensional coupling constant $\kappa$
(if necessary to infinity), such that,
under the dimensional reduction $\pr: M\to \ol{M}_0$,
\beq{2.10}
\kappa_0:=\kappa \cdot \mu^{-\frac{1}{2}}
\eeq
becomes the ${D}_0$-dimensional physical coupling constant.
If ${D}_0=4$, then ${\kappa_0}^2=8\pi G_N$,
where $G_N$ is the Newton constant.
The limit  $\kappa\to\infty$ for $\mu\to\infty$
is in particular
harmless, if $D$-dimensional gravity is given purely by curvature geometry,
without additional matter fields.
If however this geometry is coupled with finite strength
to additional (matter) fields,
one should indeed better take care to
have all internal spaces $M_i$, $i=1,\ldots,n$
compact.
If for some homogeneous space this is a priori not the case,
it often can still be achieved by factorizing this space by
an appropriate finite symmetry group.

With the total dimension $D$ , $\kappa^2$ a $D$-dimensional
gravitational constant and $\Lambda$ a $D$-dimensional
cosmological constant we consider an action of the form
\beq{2.11}
S = \frac{1}{2\kappa^2} \int_{M} d^{D}z \sq g
\{ R[g] - 2\Lambda \} + S_{\rm GHY} + S_{\Phi}+S_{\rho} .
\eeq
Here a (generalized) Gibbons-Hawking-York \cite{GH,Y} type
boundary contribution $S_{\rm GHY}$ to the action is taken
to cancel boundary terms.
Eqs.\rf{2.6} and \rf{2.7} show that $S_{\rm GHY}$ should be taken in the form
\bear{2.12}
S_{\rm GHY} &:=& \frac{1}{2\kappa^{2}} \int_{M} d^{D}z \sq g
     \{ e^{-2\ol \gamma} R_{B} \}
\nn\\
& = &\frac{1}{\kappa^2_0}
\int_{\ol{M}_0}d^{{D}_0}x
\partlx{\lambda}
\left(
e^{f}
\sqrt{|{\ol{g}^{(0)}}|} \ol{g}^{(0)\lambda\nu}
\partlx{\nu} (f+\ol{\gamma})
\right) ,
\ear
which is just a pure boundary term in form of an effective 
${D}_0$-dimensional flow through $\p \ol{M}_0$.

Also the other additional $D$-dimensional action terms depend effectively 
only on $\ol{M}_0$, like
\beq{2.13}
S_{\Phi}
=-\frac{1}{2}\int_{M} d^Dz \sqrt{|{g}|} C(\p\Phi,\p\Phi)
=-\frac{1}{2\kappa^2}\int_{M} d^Dz \sqrt{|{g}|}
C_{ab}g^{MN}\p_M\Psi^a\p_N\Psi^b ,
\eeq
generated from a metric $C$ on $k$-dimensional target space
evaluated
on a rescaled target vector field
$\Psi:=\kappa \Phi$
built from a finite number of scalar matter fields components
$\Psi^a$, $a=1,\ldots,k$, depending only on $\ol{M}_0$, 
and
\beq{2.14}
S_{\rho}
=-\int_{M} d^Dz \sqrt{|{g}|} \rho
\eeq
from a general effective matter density $\rho$
corresponding a potential on $\ol{M}_0$ which may e.g.  be chosen to 
account for the Casimir effect \cite{CW}, a Freund-Rubin monopole 
\cite{FR}, or a perfect fluid \cite{GIM,KZ}.

After dimensional reduction the action \rf{2.11} reads
\bear{2.15}
S=\frac{1}{2\kappa^2_0}
\int_{\ol{M}_0}d^{{D}_0}x
\sqrt{|{\ol{g}^{(0)}}|}
e^f
\left\{
R[\ol{g}^{(0)}]
+(\p f)(\p[f+2\ol{\gamma}])
-\sum_{i=1}^{n} d_i (\p\beta^i)^2
\right.
\nn\\
\left.
-({D}_0-2) (\p\ol{\gamma})^2
-C_{ab}(\p\Psi^a)(\p\Psi^b)
+ e^{2\ol\gamma}\left[\sum_{i=1}^{n}e^{-2\beta^i} R_i
-2\Lambda - 2\kappa^2\rho\right]
\right\} ,
\ear
where $e^f$ is a dilatonic scalar field coupling to the
$D_0$-dimensional geometry on
$\ol{M}_0$.

According to the considerations above,
due to the conformal reparametrization invariance
of the geometry on $\ol{M}_0$, we should fix a
conformal frame on $\ol{M}_0$. But then in  \rf{2.15} $\ol{\gamma}$, and
with \rf{2.5} also $f$,
is no longer independent from the vector field $\beta$, but rather
\beq{2.16}
\ol{\gamma}\equiv \ol{\gamma}[\beta]\ ,\qquad f\equiv f[\beta] .
\eeq
Then, modulo the
conformal factor $e^f$,
the dilatonic kinetic term of \rf{2.15}
takes the form
\beq{2.17}
(\p f)(\p[f+2\ol{\gamma}]) -\sum_{i=1}^{n} d_i (\p\beta^i)^2
-({D}_0-2) (\p\ol{\gamma})^2= -  G_{ij} (\p\beta^i) (\p\beta^j) ,\
\eeq
with $G_{ij}\equiv{}^{(\ol{\gamma})}\!G_{ij}$, where
\bear{2.18}
{}^{(\ol{\gamma})}\!G_{ij}&:=& {}^{(\BD)}\!G_{ij}
 -({D}_0-2)({D}_0-1) \, \frac{\p\ol{\gamma}}{\p\beta^i}
\frac{\p\ol{\gamma}}{\p\beta^j}
-2 (D_0-1) d_{(i} \, \frac{\p\ol{\gamma}}{\p\beta^{j)} } ,\
\\
\label{2.19}
&& {}^{(\BD)}\!G_{ij} := \delta_{ij} d_i - d_i d_j .
\ear
For $D_0\neq 2$, we can write equivalently
$G_{ij}\equiv{}^{(f)}\!G_{ij}$, where
\bear{2.20}
{}^{(f)}\!G_{ij} &:=& {}^{(\E)}\!G_{ij}
-\frac{{D}_0-1}{{D}_0-2} \, \frac{\p f}{\p\beta^i}
\frac{\p f}{\p\beta^j} ,\
\\
\label{2.21}
&& {}^{(\E)}\!G_{ij} :=  \delta_{ij} d_i + \frac{d_i d_j}{{D}_0-2} .
\ear
For ${D}_0=1$, $G_{ij}= {}^{(\E)}\!G_{ij}={}^{(\BD)}\!G_{ij}$ is
independent of $\ol{\gamma}$ and $f$.
Note that the metrics \rf{2.19} and \rf{2.21} (with $D_{0}\neq 2$)
may be diagonalized by appropriate homogeneous linear minisuperspace 
coordinate transformations (see e.g.  \cite{RZ,GZ,Ra0})
to $({\mp}({\pm})^{\delta_{1{D}_0}})^{\delta_{1i}} \delta_{ij}$ 
respectively.

After gauging $\ol{\gamma}$, setting $m:=\kappa^{-2}_0$, 
\rf{2.15} yields a $\sigma$-model 
in the form 
\bear{Sgamma0}
{}^{(\ol \gamma)}\!S=\int_{\ol M_0} d^{{D}_0}x \sqrt{|\ol g^{(0)}|} 
{}^{(\ol \gamma)}\!N^{{D}_0} &\!  &\!  \phi(\beta) \
\left\{
\frac{m}{2} {}^{(\ol \gamma)}\!N^{-2} 
\left[ R[\ol g^{(0)}] - {}^{(\ol \gamma)}\!G_{ij}(\p\beta^i)(\p\beta^j)
-C_{ab}(\p\Phi^a)(\p\Phi^b)
\right]
\right. 
\nn\\
&\! &\! \qquad \qquad 
\left.
- {}^{(\BD)}\!V(\beta)
\right\}\ ,
\\
\label{Vgamma0}
 \mbox{\rm where}\qquad  
{}^{(\BD)}\!V(\beta) \ :=&\! &\! m\left[
\Lambda+\kappa^2\rho-\frac{1}{2} \sum_{i=1}^{n} R[g^{(i)}]
e^{-2\beta^i} \right] \ , 
\\
\label{Ngamma0}
{}^{(\ol \gamma)}\!N \ :=&\! &\! e^{\ol \gamma}\ . 
\ear
Note that, the potential \rf{Vgamma0} 
and the conformal factor $\phi(\beta):=\prod\nolimits_{i=1}^{n}e^{d_i\beta^i}$
are gauge invariant.

Analogously, the $\sigma$-model action from \rf{2.15} gauging $f$ can also be 
written as
\bear{Sf0}
{}^{(f)}\!S=\int_{\ol M_0} d^{{D}_0}x \sqrt{|\ol g^{(0)}|} 
{}^{(f)}\!N^{{D}_0}
 &\! &\! 
\left\{\frac{m}{2}
{}^{(f)}\!N^{-2} 
\left[ R[\ol g^{(0)}] - {}^{(f)}\!G_{ij}(\p\beta^i)(\p\beta^j) 
-C_{ab}(\p\Phi^a)(\p\Phi^b)
\right]
\right. 
\nn\\
&\! &\! \qquad \qquad 
\left.
- {}^{(\E)}\!V(\beta)
\right\} \ , 
\\
\label{Vf0}
{}^{(\E)}\!V(\beta)\ := &\! &\!  
m \Omega^2
\left[
\Lambda+\kappa^2\rho-\frac{1}{2} \sum_{i=1}^{n} R[g^{(i)}]
e^{-2\beta^i} \right]\  ,
\\
\label{Nf0}
{}^{(f)}\!N\ := &\! &\! e^{\frac{f}{{D}_0-2}}\ , 
\ear
where the function $\Omega$ on $\ol{M}_0$ is defined as 
\beq{Omega}
\Omega:=\phi^{ \frac{1}{2-{D}_0} }\ .
\eeq 
Note that, with $\Omega$ also the potential \rf{Vf0} 
is gauge invariant,
and the dilatonic target-space, 
though not even conformally flat in general,
is flat for constant $f$.

In fact, Eqs. \rf{Sgamma0}-\rf{Ngamma0} and \rf{Sf0}-\rf{Nf0} show that there 
are at least two special frames.  

The first one corresponds to the gauge $\ol\gamma \mustbe 0$.  
In this case 
${}^{(\ol \gamma)}\!N=1$, the minisuperspace metric
\rf{2.18} reduces to the Minkowskian \rf{2.19},
the dilatonic scalar field becomes proportional to the internal space volume, 
$e^{f[\beta]}=\phi(\beta)=\prod\nolimits_{i=1}^{n}e^{d_i\beta^i}$, and
\rf{Sgamma0} describes a generalized $\sigma$-model with conformally 
Minkowskian target space.  The Minkowskian signature implies a negative 
sign in the dilatonic kinetic term.  This frame is usually called the 
Brans-Dicke one, because $\phi=e^f$ here plays the role of the 
Brans-Dicke scalar field.  Our effective theory following from 
multidimensional cosmology \cite{RZ} takes a generalized Brans-Dicke 
form.

The second distinguished frame corresponds to the gauge $f \mustbe 0$,
where $\ol\gamma =\frac{1}{2 - D_0}\sum_{i=1}^n d_i\beta^i $ 
is well-defined only for $D_0\neq 2$.  In this case 
${}^{(f)}\!N=1$, the minisuperspace metric
\rf{2.20} reduces to the Euclidean \rf{2.21}, and \rf{Sf0} describes a 
self-gravitating $\sigma$-model with Euclidean target space.  Hence all 
dilatonic kinetic terms have positive signs.  This frame is usually called 
the Einstein one, because it describes an effective $D_0$-dimensional 
Einstein theory with additional minimally coupled scalar fields.  
For multidimensional geometries with $D_0 = 2$ the Einstein frame 
fails to exist, which reflects the well-known fact that two-dimensional 
Einstein equations are trivially satisfied without implying any dynamics.

For ${D}_0=1$, the action of both \rf{Sgamma0} and \rf{Sf0} was shown in
\cite{Ra0} (and previously in \cite{Ra1,Ra2}) to take the form
of a classical particle motion on minisuperspace, whence different frames 
correspond are just related by a time reparametrization.
More generally, for ${D}_0\neq 2$ and $(\ol M_0,\ol g{(0)})$
a vacuum space-time, 
the $\sigma$-model \rf{Sf0} with the gauge $f \mustbe 0$ 
describes the dynamics of a massive
$({D}_0-1)$-brane within a potential \rf{Vf0}
on its target minisuperspace. 

Before concluding this chapter, let
us point out that besides the Brans-Dicke and the Einstein gauge,
which are the main topic of this paper,  
there might be further gauges of interest 
for particular physical features.

{}From \rf{2.15} we see that, 
there exists another similarly distinguished frame,
namely the one corresponding to the gauge 
$f\mustbe -2\ol \gamma=\frac{2}{D_0}\sum_{i=1}^{n}d_i\beta^i$,
in which, as for the Einstein frame, the kinetic term 
$(\p f)(\p[f+2\ol{\gamma}])$ carrying the anomalous sign vanishes,
whence the target minisuperspace carries a true (not just a pseudo) metric
corresponding to a non-negative kinetic contribution to the action.
In this gauge the potential terms decouple
{}from the dilatonic field $f$, although the latter 
still couples to the kinetic terms.

Of course the choice of any a priori prescribed action
strongly affects the ``natural'' choice of frame.
For different theories we can introduce different
"natural" gauges.

For example starting from an $D$-dimensional effective string action
which includes besides the dilaton also a massless axion
there is a so-called ``axion'' gauge \cite{CEW}
which decouples the axion
{}from the dilaton field.


We conclude by emphasizing again that for theories \rf{1.1} with action 
of the type \rf{2.11} there exist compelling physical arguments in favor of 
the Einstein frame.  
Therefore we will now investigate how to generate solutions 
in this frame.

\section{\bf Generating solutions in the Einstein frame}
\label{Sect. 3}
\setcounter{equation}{0}
In the following we denote the external space-time metric $\ol{g}^{(0)}$
in the Brans-Dicke frame with $\ol{\gamma}\mustbe 0$
as $\overline{g}^{(\BD)}$
and in the Einstein frame with $f\mustbe 0$
as $\overline{g}^{(\E)}$.
It can be easily seen that they are connected with each other
by a conformal transformation
\beq{3.1}
\overline{g}^{(\E)}
\mapsto
\overline{g}^{(\BD)} = \Omega^2 \overline{g}^{(\E)} 
\eeq
with $\Omega$ from \rf{Omega}.

Let us now consider the space time foliation $\ol{M}_0=\R\times M_0$
where $g^{(0)}$ is a smooth homogeneous metric on $M_0$.
Under any projection $\pr_0: \overline{M_{0}} \to \R$
a consistent pullback
of the metric
$- e^{2\gamma(\tau)} d\tau\otimes d\tau$
{}from $\tau\in \R$ to $x\in\pr^{-1}_0\{\tau\}\subset \ol{M}_0$
is given by
\beq{3.3}
\overline{g}^{(\BD)}(x)
:= - e^{2\gamma(\tau)} d\tau\otimes d\tau
+ e^{2\beta^0(x)} {g}^{(0)} .
\eeq
For spatially (metrically-)homogeneous cosmological models
as considered below
all scale factors $a_i:=e^{\beta^i}$, $i=0,\ldots,n$
depend only on $\tau\in\R$.

With \rf{3.3} and \rf{3.1}, Eq.
\rf{1.2} reads
\bear{3.4}
g
&=& - e^{2\gamma(\tau)} d\tau\otimes d\tau + a_{0}^2 g^{(0)}
+\sum_{i=1}^{n} e^{2\beta^i} g^{(i)}
\nn\\
&=& -  dt_{\BD} \otimes dt_{\BD} + a^2_{\BD} g^{(0)}
+\sum_{i=1}^{n} e^{2\beta^i} g^{(i)}
\nn\\
&=&  -  \Omega^2 dt_{\E} \otimes dt_{\E} + \Omega^2 a^2_{\E} g^{(0)}
+\sum_{i=1}^{n} e^{2\beta^i} g^{(i)} ,
\ear
where $a_0:= a_{\BD}$ and $a_{\E}$ are the external space scale factor
functions depending respectively on the cosmic synchronous time
$t_{\BD}$ and $t_{\E}$ in the Brans-Dicke and Einstein frame.
With \rf{Omega} the latter is related to the former by
\beq{3.5}
a_{\E}
= \Omega^{-1} a_{\BD}
=\left(
\prod_{i=1}^n e^{d_i\beta^i}
\right)^{ \frac{1}{{D}_0-2} }
a_{\BD} ,
\eeq
and the cosmic time of the Einstein frame is given by
\beq{3.6}
|dt_{\E}|
=|\Omega^{-1}e^{\gamma}d\tau|
=\left|\left(
\prod_{i=1}^n e^{d_i\beta^i}
\right)^{ \frac{1}{{D}_0-2} }
dt_{\BD} \right|\ .
\eeq
As a consequence of the arguments mentioned in the introduction,
$t_{\E}$ will be considered below as the physical time.
The presently best known (spatially homogeneous) cosmological solutions
with a metric structure given by \rf{1.2} and \rf{3.3}
were found in the Brans-Dicke frame
(see e.g. \cite{IM,GIM,KZ,KRZ} and an extensive list of references there).
Most of them are described most simply within one of the following
two systems of target space coordinates.
We set
\bear{3.7}
q:=\sqrt{\frac{D-1}{D-2}}\ ,
&\qquad &
p:=\sqrt{\frac{{d}_0-1}{d_0}}\ .
\ear
With $\Sigma_k=\sum_{i=k}^{n}d_i$,
the first coordinate system \cite{IMZ} is related to $\beta^i$, $i=0,\ldots,n$,
as
\bear{3.8}
z^0&:=&q^{-1}\sum_{j=0}^{n}d_j\beta^j\ ,
\nn\\
z^i&:=&
{\left[\left.d_{i-1}\right/\Sigma_{i-1}\Sigma_{i}\right]}^{1/2}
\sum_{j=i}^{n}
d_j\left(\beta^j-\beta^{i-1}\right)\ ,\quad i=1,\ldots,n\ ,
\ear
and the second one \cite{Z} as
\bear{3.9}
v^0&:=&p^{-1}(\sum_{j=0}^{n}d_j\beta^j - \beta^0) \ ,
\nn\\
v^1 &:=& p^{-1}[(D-2)/d_0\Sigma_1]^{1/2}
\sum_{j=1}^n d_j\beta^{j} ,
\nn\\
v^i&:=&
{\left[\left.d_{i-1}\right/\Sigma_{i-1}\Sigma_{i}\right]}^{1/2}
\sum_{j=i}^{n}
d_j\left(\beta^j-\beta^{i-1}\right)\ ,\quad i=2,\ldots,n\ ,
\ear
In both of this minisuperspace coordinates
the  target space Minkowski metric $G$ is given in form of
the standard diagonal matrix $G_{ij}:=(-)^{\delta_{0i}}\delta_{ij}$.
The two coordinates are related by a Lorentz boost in the
$(01)$-plane.

In coordinates \rf{3.8} some known solutions
(see e.g. \cite{KRZ,BIMZ,Z2}) take the form
\beq{3.10}
a_i = A_i ( e^{q z^0})^{\frac{1}{D-1}} e^{\alpha^i \tau} ,
\quad i=0,\ldots,n ,
\eeq
where parameters $\alpha^i$ satisfy conditions
\bear{3.11}
\sum_{i=0}^{n}d_i\alpha^i & = & 0 ,
\\\nn
\sum_{i=0}^{n}d_i(\alpha^i)^2 & = & 2\epsilon
\ear
and $\epsilon$ is a non-negative parameter.

In coordinates \rf{3.9} some known solutions
(see e.g. \cite{KZ,BZ1}) take the form
\bear{3.12}
a_0 &=& A_0 ( e^{p v^0})^{\frac{1}{d_0-1}} e^{\alpha^0 \tau} ,
\\\nn
a_i &=& A_i  e^{\alpha^i \tau} ,
\quad i=1,\ldots,n ,
\ear
where parameters $\alpha^i$ satisfy conditions
\bear{3.13}
\sum_{i=0}^{n}d_i\alpha^i & = & \alpha^0 ,
\\\nn
\sum_{i=0}^{n}d_i(\alpha^i)^2 & = & (\alpha^0)^2 + 2\epsilon
\ear
and $\epsilon$ is a non-negative parameter.

Explicit expressions for functions $z^0\equiv z^0(\tau )$ 
and $v^0\equiv v^0(\tau )$ depend on the details of the particular models.

Solutions of the form \rf{3.10} and \rf{3.12} were found in the harmonic time 
gauge $\gamma\mustbe\sum_{j=0}^{n}d_j\beta^j$, where $\tau$ is the 
harmonic time of the Brans-Dicke frame.  Equation \rf{3.8} shows that the 
coordinate $z^0$ is related to the dynamical part of the total spatial 
volume in the BD frame: $v:=e^{q z^0}=\prod_{i=0}^{n}{a_i}^{d_i}$.

Relations  \rf{3.5} and \rf{3.9}
between the different minisuperspace coordinates
imply that
\beq{3.14}
(d_0-1) \beta^0_E
=(d_0-1)\beta^0 + \sum_{j=1}^{n}d_j\beta^j
=p v^0 \ ,
\eeq
which shows that the coordinate $v^0$ is proportional
to the logarithmic scale factor of external space
in the Einstein frame: $a_E := e^{\beta ^0_E}$.

Thus target space coordinates $z$ have the most natural interpretation
in the Brans-Dicke frame,
whereas target space coordinates $v$ are better adapted to the
Einstein frame.

Via \rf{3.14} synchronous time in the Einstein frame
is related to harmonic time $\tau$ in the Brans-Dicke frame
by integration of \rf{3.6} with integration constant $c$ to
\beq{3.15}
|t_{\E}|+c
=\int \left( e^{p v^0}\right)^{{d_0}/{d_0-1}} {d\tau}
=\int a^{d_0}_{\E}{d\tau}\ .
\eeq
Thus the physical metric of external space-time reads
\beq{3.16}
{g}^{(\E)}= - a_E^{2 d_0} d\tau\otimes d\tau
+ a^2_{\E} {g}^{(0)} ,
\eeq
where for solutions \rf{3.10}
\beq{3.17}
a_{\E} =
\left[
\frac{ ( e^{q z^0})^{\frac{1}{{q}^2}} }{ A_0 e^{\alpha^0 \tau} }
\right]^{\frac {1}{d_0-1}} ,
\eeq
and for solutions \rf{3.12}
\beq{3.18}
a_{\E} =
 \left( e^{p v^0}\right)^{ \frac{1}{d_0-1} } .
\eeq
Expressions for the internal scale factors are not
affected.
In Eqs. \rf{3.16} to  \rf{3.18} the time $\tau$ is the harmonic
one from the Brans-Dicke frame.
The transformation to synchronous time in the
Einstein frame is provided by Eq. \rf{3.15}.
Once $z^0$
or $v^0$ is known as a function of $\tau$,
explicit expressions can be given.
However these functions depends on the concrete form
of the cosmological model
(see \cite{IM}-\cite{KRZ}, \cite{Z}-\cite{BZ1}).

Above we obtained a general prescription for
the generation of solutions in the Einstein frame
{}from already known ones in the Brans-Dicke frame.
It can easily be seen that the behavior of the solutions
in both of these frames is quite different.
Let us demonstrate this explicitly
by the example of a generalized Kasner solution.

\section{\bf Solutions in original form}
\label{Sect. 4}
\setcounter{equation}{0}
Let $t:= t_{\BD}$ be the synchronous time of the Brans-Dicke frame,
and $\dot x$ denote the derivative of $x$ with respect to $t$.

The well-known Kasner solution \cite{Ka} describes
a $4$-dimensional anisotropic space-time with the metric
\beq{4.1}
g = -  dt\otimes dt + \sum_{i=1}^{3} t^{2p_i} dx^i\otimes dx^i
\eeq
where the $p_i$ are constants satisfying
\beq{4.2}
\sum_{i=1}^{3} {p_i} =\sum_{i=1}^{3} {(p_i)}^2 = 1 .
\eeq
It is clear that a multidimensional generalization
of this solution is possible for a manifold \rf{1.1}
with  Ricci flat factor spaces $(M_i,g_i)$, $i=0,\ldots,n$.
Particular solutions generalizing \rf{4.1} with \rf{4.2}
were obtained in many papers \cite{ChD}-\cite{Ok}.
More general solutions for an arbitrary number
of $d_i$-dimensional tori were found in \cite{Iv}
and generalized to the case of a free minimally coupled
scalar field $\Phi$ in \cite{BlZ}.
In the latter case there are two classes of solutions.

A first class represents namely Kasner-like solutions.
None of these is contained in the hypersurface
\beq{4.3}
\sum_{i=0}^{n} d_i\dot\beta^i = 0
\eeq
of constant spatial volume.
With $c$ and $a_{(0)i}$, $i=0,\ldots,n$ integration constants,
in the Brans-Dicke synchronous time gauge such a solution reads
\bear{4.4}
a_i&=&a_{(0)i} t^{\ol\alpha^i} ,\quad i=0,\ldots,n ,
\\\label{4.4a}
\Phi&=&\ln t^{\ol \alpha^{n+1}} + c ,
\ear
where   the $\ol \alpha^i$
fulfill the conditions
\bear{4.5}
\sum_{i=0}^{n} d_i{\ol\alpha^i}&=&1 ,
\\\nn
\sum_{i=0}^{n} d_i{(\ol\alpha^i)}^2&=&1 - ({\ol \alpha^{n+1}})^2 .
\ear
Without an additional non-trivial scalar field $\Phi$,
i.e. for ${\ol \alpha^{n+1}}=0$,
these conditions become analogous to \rf{4.2}
\beq{4.6}
\sum_{i=0}^{n} d_i{\ol\alpha^i}
=\sum_{i=0}^{n} d_i{(\ol\alpha^i)}^2=1 .
\eeq
Solution \rf{4.4} describes a universe with
increasing total spatial volume
\beq{4.7}
v \sim \prod_{i=0}^{n} {a_i}^{d_i} \sim t
\eeq
and decreasing Hubble parameter for each factor space
\beq{4.8}
h_i := \frac{1}{a_i}\frac{da_i}{dt} = \frac{\ol\alpha^i}{t} ,
\quad i=0,\ldots,n .
\eeq

In the case of imaginary scalar field ($(\ol \alpha^{n+1})^2 < 0$) 
factor spaces with $\ol\alpha^i >1$ undergo a power law inflation.  
The absence of a non-trivial scalar field, i.e.  $\Phi \equiv 0$, implies
(except for $d_0=\ol \alpha^0 =1$, $\ol \alpha^i =0$, $i=1,\ldots,n$) 
that $|\ol \alpha^i | < 1$ for $i=0,\ldots,n$.  
In \cite{Le} it was shown that after a transformation $t \to t_0 - t $ 
(reversing the arrow of time) factor spaces with $\ol\alpha^i < 0$ can 
be interpreted as inflationary universes with scale factors $a_i
\sim (t_0-t)^{\ol\alpha^i}$ with $\ol\alpha^i < 0$ growing at an accelerated 
rate $\ddot a_i / a_i > 0$.

A second (more special) class of solutions
is confined to the hyperplane \rf{4.3} in momentum space.
In this case (in the Brans-Dicke frame)
harmonic and synchronous time coordinates
coincide and solutions read
\bear{4.9}
a_i&=&a_{(0)i} e^{b^i t} ,\quad i=0,\ldots,n ,
\\\label{4.9a}
\Phi&=&b^{n+1} t + c ,
\ear
where $c$ is a integration constant and the constants $b^i$ satisfy
\bear{4.10}
\sum_{i=0}^{n} d_i{b^i}&=&0 ,
\\\nn
\sum_{i=0}^{n} d_i{(b^i)}^2 + ({b^{n+1}})^2&=&0 .
\ear
The latter relation shows that these solutions
are only possible if $\Phi$ is an imaginary scalar field
with $(b^{n+1})^2<0$.

The inflationary solution \rf{4.9} describes
a universe with constant total spatial volume
\beq{4.11}
v \sim \prod_{i=0}^{n} {a_i}^{d_i} = \prod_{i=0}^{n} a^{d_i}_{(0)i} \ ,
\eeq
and a nonzero but constant Hubble parameter
\beq{4.12}
h_i = \frac{1}{a_i}\frac{da_i}{dt} = b^i\ ,
\quad i=0,\ldots,n\ ,
\eeq
for each factor space.
This is a particular case
of a steady state universe
where stationarity of matter energy density in the whole universe is
maintained due to redistribution
of matter between contracting and expanding parts
(factor spaces) of the universe
(matter density in the whole universe is constant due to
the constant volume).
This is unlike the original steady-state
theory \cite{BG}, where a continuous creation of matter is required
in order to stabilize matter density, which then
necessitates a deviation from Einstein  theory.
In \cite{Ra0} the inflationary solution was generalized for the case of
a $\sigma$-model with $k$-dimensional target vectors $\Phi$
rather than a single scalar field.

\section{\bf Solutions in the Einstein frame}
\label{Sect. 5}
\setcounter{equation}{0}
Let us now transform the 
solutions \rf{4.4}, \rf{4.4a} 
and   \rf{4.9},\rf{4.9a} above
to the Einstein frame, using the general prescription from Sec. \ref{Sect. 3}.

We first consider the Kasner-like solution \rf{4.4},
where \rf{Omega} determines the conformal factor as
\beq{5.1}
\Omega^{-1}=\left( \prod_{i=1}^{n} e^{d_i\beta^i}\right)^{\frac{1}{ D_0-2}}
=C_1  t^{(1-d_0\ol\alpha^0)/(d_0-1)}
\eeq
with
\beq{5.2}
C_1 := \left( \prod_{i=1}^{n} a_{\!(0)i}^{\ d_i} \right)^{\frac{1}{D_0-2}} .
\eeq
As it was noted above, the conformal transformation to
the Einstein frame does not exist for $ D_0=2$ ($d_0=1$).
In the special case of $\ol\alpha^0=\frac{1}{d_0}$ the conformal factor
$\Omega$ is constant, and both frames represent the same 
connection, hence the same geometry.
\footnote{Here is meant the geometry as given by the connection.
Locally at $x\in \ol M_0$ this is just the
$\End(T_x\ol M_0)$-valued Riemannian curvature $2$-form.}
Even in this case, \rf{5.2} is still divergent  for $d_0=1$.

The external space scale factor in the Einstein frame
(physical scale factor of the external space) is defined
by formula \rf{3.5} which for \rf{5.1} reads
\beq{5.3}
a_{\E} = \Omega^{-1} a_{\BD} =
\ol a_{0} t^{(1-\ol\alpha^0)/(d_0-1)} ,
\eeq
where $\ol a_{0}:=C_1 a_{(0)0}$.
At $\ol\alpha^0=\frac{1}{d_0}$ the (external space)
scale factor $a_{\E} = \ol a_{0} t^{\ol \alpha^0}
\sim a_{\BD}$
has the same behavior in both frames which is just what
one expects for constant $\Omega$.

So the physical metric of the external space-time
reads
\beq{5.4}
\ol{g}^{(\E)}
= -\Omega^{-2} dt \otimes dt + a^2_{\E} {g}^{(0)}
= - dt_{\E} \otimes dt_{\E}  + a^2_{\E} {g}^{(0)} ,
\eeq
where $\Omega^{-1}$ and $a_{\E}$ are given by equations
\rf{5.1} and \rf{5.3}
respectively, and $t$ is given synchronous time in the Brans-Dicke
frame connected with synchronous time in the Einstein frame
via \rf{3.6}.
Putting the integration constant to zero we obtain
\beq{5.5}
t=C_2 t^{(d_0-1)/d_0(1-\ol\alpha^0)}_{\E} ,
\eeq
where $C_2=\left[C^{-1}_1\frac{1-\ol\alpha^0}{{d_0-1}}d_0
\right]^{(d_0-1)/d_0(1-\ol\alpha^0)}$.
The value $\ol\alpha^0=1$ is a singular one.
It can be seen from
\rf{4.5} that $|\ol\alpha^i|<1$, $i=0,\ldots,n+1$
when the scalar field is real.
The value $\ol\alpha^0=1$ 
may appear only in the case of
an imaginary scalar field.
\rf{5.3} shows that, in this case $a_{\E}$ is a constant.
In the case $\ol\alpha^0\neq 1$ the generalized Kasner-like solutions 
in the Einstein frame take the form
\bear{5.6}
a_{i,\E} &=& \tilde a_{i} {t_{\E}}^{\tilde\alpha^i} ,
\quad i = 0,\ldots,n,
\\\label{5.6a}
\Phi&=&{\tilde\alpha^{n+1}} \ln {t_{\E}} +  c .
\ear
Here and in the following
$a_{0,\E}:=a_{E}(t_{E})$, $a_{i,\E}:=a_{i}(t_{E})$, $i=1,\ldots,n$,  
are given as functions of $t_{\E}$, while
$\tilde a_{i}$, $i=0,\ldots,n$,  and $c$ are constants.
In \rf{5.6} and \rf{5.6a} the powers $\tilde \alpha^i$ are defined as
\bear{5.7}
\tilde\alpha^{0}&:=&\frac{1}{d_0}
\\\nn
\tilde\alpha^{i}&:=&\frac{d_0-1}{d_0(1-\ol\alpha^0)} \ol\alpha^i ,
\quad i = 1,\ldots,n+1,
\ear
with $\ol\alpha^i$, $i=0,\ldots,n+1$, satisfying relations \rf{4.5}.  
Hence in 
contrast to \rf{4.4} there is no freedom in the choice of the power 
$\tilde\alpha^{0}$.  For example at $d_0=3$ one obtains a physical external 
space scale factor $a_{\E}={t_\E}^{1/3}$, i.e.  the external space 
$(M_0,g_0)$ behaves like a Friedmann universe filled with ultra stiff 
matter (which is equivalent to a minimally coupled scalar field).

Let us emphasize here once more that in the present approach the physical 
theory is modeled as a $ D_0$-dimensional effective action with the 
space-time metric \rf{5.4} in the Einstein frame ($f=0$).  All internal 
spaces are displayed in the external space-time as scalar fields, leading 
to a $ D_0$-dimensional self-gravitating $\sigma$-model with 
self-interaction \cite{RZ}.


Let us transform now the inflationary solution \rf{4.9}
to the Einstein frame.
For this solution the conformal factor and the external space
scale factor read
\bear{5.15}
\Omega^{-1} &=& C_1 \exp \left(-\frac{d_0 b^0}{d_0-1} t\right) ,
\\
a_{\E} &=& \ol a_{0} \exp \left(-\frac{b^0}{d_0-1} t\right) ,
\ear
where $C_1$ is defined by \rf{5.2} and
$\ol a_{0}=C_1 a_{(0)0}$.
Note that the conformal transformation \rf{5.15} breaks down
for $ D_0=2$ ($d_0=1$). This happens even in the special case 
of $b^{0}=0$. For the latter, $\Omega$ is constant, 
whence the connection and its geometry represented by both frames
are the same. Here, the external space is static in both of them.

For $b^{0} \neq 0$, synchronous times in the Brans-Dicke
and Einstein frames are related by
\beq{5.17}
t = \frac{d_0-1}{d_0 b^0} \ln ( C_2 {t_{\E}}^{-1}) ,
\eeq
where (taking a relative minus sign in \rf{3.6})
$C_2=C_1\frac{d_0-1}{d_0 b^0}$.
Thus in the Einstein frame scale factors have power-law behavior
\beq{5.18}
a_{i,\E}=\tilde a_{i} { t_{\E} }^{\tilde\alpha^i} ,
\quad i=0, \ldots, n ,
\eeq
with
\bear{5.19}
\tilde\alpha^{(0)}&:=&\frac{1}{d_0}
\\\nn
\tilde\alpha^{(i)}&:=&-\frac{d_0-1}{d_0}\frac{b^i}{b^0}
\quad i= 1,\ldots, n .
\ear
Similar as for the Kasner-like solution, the inflationary solution transformed 
to the Einstein frame has no freedom in choice of the power $\tilde 
\alpha^{(0)}$.  The external space scale factor behaves as $a_{0,\E} \sim 
t_{\E}^{1/d_0}$ (compare also \rf{5.6} and \rf{5.7}).  The scalar field 
reads
\beq{5.21}
\Phi = \tilde \alpha^{n+1} \ln t_{\E} + c ,
\quad \tilde \alpha^{n+1} := -\frac{d_0-1}{d_0}\frac{b^{n+1}}{b^0} .
\eeq
Using \rf{4.10} we obtain the sum rules
\bear{5.22}
\sum_{i=0}^{n}d_i\tilde \alpha^i & = & d_0 ,
\\\nn
(\tilde \alpha^{n+1})^2 & = &  {2-d_0}
- \sum_{i=0}^{n}d_i(\tilde \alpha^i)^2 < 0 ,
\ear
whence the scalar field is imaginary.

The main lesson we learned in this section
is the following:
The dynamical behavior of scale factors and scalar fields
strongly depends on the choice of the frame.
For example in the case of solutions
originating from the Kasner and inflationary solutions
of Sec. \ref{Sect. 4} the external space scale factor
in the Einstein frame behaves as $t_{\E}^{1/d_0}$
(except for the cases $\ol\alpha^0=1$ and $b^0=0$ where $a_{\E}$ is
a constant).
In this case there is no inflation of the external space,
neither exponential
nor power law (with power larger than $1$).
In contrast to the conclusions drawn in \cite{Le}
for the Kasner solutions in the Brans-Dicke frame,
inversion of arrow of time $t_{\E}$ in the Einstein frame
does not lead to inflation of the external space because of
power $1/d_0 > 0$.

\section{\bf Conclusions }
\setcounter{equation}{0}
We started from a higher-dimensional cosmological model
based on a smooth manifold of topology \rf{1.1}
with a multidimensional geometry given by a metric ansatz \rf{1.2}.

Then, an Einstein theory in higher dimension $D$
can be reduced to an effective model in lower dimension ${D}_0$.
This is a (generalized) $\sigma$-model
with  conformally flat target space.
With a purely geometrical dilaton field $f$,
it provides a natural generalization for
the well-known Brans-Dicke theory.

In the Introduction we gave several reasons which suggest
that Einstein frame with $f=0$ should be the
preferred frame for  a more direct physical interpretation
of the model under consideration.
This necessitates that, before a physical
interpretation can be given,
solutions previously obtained  in the
Brans-Dicke frame should first be transformed
to the Einstein frame.

Typical solutions for considered models in Brans-Dicke frame
have a general  structure described either
by \rf{3.10} or \rf{3.12}.
For solutions of this type
the transformation to Einstein frame is given
by \rf{3.17} and \rf{3.18} respectively.
The qualitative difference induced by the distinct functions
$z^0$ and $v^0$ respectively
necessitates a separate treatment of
these two classes.
In any case, solutions to a given model
in the Einstein frame show a  different qualitative behavior
{}from the corresponding solution in the Brans-Dicke frame.

We demonstrated this explicitly on the example
of the generalized Kasner solution \rf{4.4}
(and exceptional inflationary solutions \rf{4.9}).
With respect to the proper time in Einstein frame,
the external space scale factor $a_{0,\E}$
has a surprisingly simple and definite
root law behavior
$a_{0,\E} \sim t^{1/{d}_0}_{\E}$
(except for the case of an exotic imaginary scalar field
where $a_{0,\E}$ may be constant).
Hence this model  does not admit inflation
of the external space
in Einstein frame.
This contrasts investigations \cite{Le} performed in 
the Brans-Dicke frame.

Similarly the transformation of all other known solutions
can give rise to new surprising results.

\nl\nl
{\Large {\bf Acknowledgments}}
\nl\nl
We are grateful for useful discussions with
L.  Garay, D.-E.  Liebscher, L.  Sokolowski, and D.  Wands on the physical 
meaning of different gauges of the conformal frame.

This work was partially supported by Deutsche Forschungsgemeinschaft, 
in particular by the grants  436 UKR 113/34 (A. Z.)
and 436 RUS 113/7 (M. R.).  
A. Z. thanks the Gravitationsprojekt  
at the Department of Mathematical Physics (Prof.  H.  Baumg\"artel) 
at Universit\"at Potsdam 
and the Institut f\"ur Theoretische Physik (Prof. H. Kleinert) 
at FU Berlin for their hospitality.
M. R. is also grateful to Profs. H. Baumg\"artel (Univ. Potsdam) 
and A. Ashtekar (Center for Gravitational Physics and Geometry, 
Penn. State) for their scientific support.


\begin{thebibliography}{99}
\bibitem{SV}
A. Strominger and C. Vafa,
{Phys. Lett. B} {\bf 379}, 99 (1996).
\bibitem{Du}
M. J. Duff,
{Int. J. Mod. Phys. A} {\bf 11}, 5623 (1996).
\bibitem{SchSch}
J. Scherk and J. H. Schwarz,
{Phys. Lett. B} {\bf 57}, 463 (1975).
\bibitem{FR}
P. G. O. Freund and M. A. Rubin,
{Phys. Lett. B} {\bf 97}, 233 (1980).
\bibitem{SSt}
A. Salam and J. Strathdee,
{Ann. Phys.(NY)} {\bf 141}, 316 (1982).
\bibitem{DNP}
M. J. Duff, B. E. W. Nilson, and C. N. Pope,
{Phys. Rep.} {\bf 130}, 1 (1986).
\bibitem{Di}
R. Dick,
{Gen. Rel. Grav.} {\bf 30}, 435 (1998).
\bibitem{MaSo}
G. Magnano and L. M. Soko\l owski,
{Phys. Rev. D} {\bf 50}, 5039 (1994).
\bibitem{RZ}
M. Rainer and A. Zhuk,
{Phys. Rev. D}  {\bf 54}, 6186 (1996).
\bibitem{Cho}
Y. M. Cho,
{Phys. Rev. Lett.} {\bf 68}, 3133 (1992).
\bibitem{Mar}
W. J. Marciano,
{Phys. Rev. Lett.} {\bf 52}, 489 (1984).
\bibitem{KPW}
E. W. Kolb, M. J. Perry, and T. P. Walker,
{\it Phys. Rev. D} {\bf 33}, 869 (1986).
\bibitem{GGB}
L. J. Garay and J. Garcia-Bellido,
{Nucl. Phys. B} {\bf 400}, 416 (1993).
\bibitem{GSZ}
U. G\"unther, A. Starobinsky, and A. Zhuk,
Gravitational excitons and matter from extra dimensions,
{
Preprint} (1998).
\bibitem{Ma}
K. I Maeda,
{Class. Quant. Grav.} {\bf 3}, 651 (1986).
\bibitem{BM}
K. A. Bronnikov and V. N. Melnikov,
{Ann. Phys. (NY)} {\bf 239}, 40 (1995).
\bibitem{GZ}
U. G\"unther and A. Zhuk,
{Phys. Rev. D} {\bf 56}, 6391 (1997).
\bibitem{CSY}
T. Chiba, N. Sugiyama, and J. Yokoyama,
Constraints on scalar-tensor theories of gravity from density
perturbations in inflationary cosmology,
{
Preprint} gr-qc/9708030 .
\bibitem{MR}
S. Malik and D. Rai Chaudhuri,
{Phys. Rev. D} {\bf 56}, 625 (1997).
\bibitem{SSM}
M. Saijo, H. Shinkai, and K.I Maeda,
{Phys. Rev. D} {\bf 56}, 785 (1997).
\bibitem{IM}
V. D. Ivashchuk and V. N. Melnikov,
{Gravitation \& Cosmology} {\bf 1}, 133 (1995).
\bibitem{GIM}
V. R. Gavrilov, V. D. Ivashchuk, and V. N. Melnikov,
{J. Math. Phys} {\bf 36}, 5829 (1995).
\bibitem{KZ}
U. Kasper and A. Zhuk,
{Gen. Rel. Grav.} {\bf 28}, 1269 (1996).
\bibitem{KRZ}
U. Kasper, M. Rainer and A. Zhuk,
{Gen. Rel. Grav.} {\bf 29}, 1123 (1997).
\bibitem{GH}
G. W. Gibbons and S. W. Hawking,
{Phys. Rev. D} {\bf 15}, 2752 (1977).
\bibitem{Y}
J. W. York,
{Phys. Rev. Lett.} {\bf 28}, 1082 (1972);
{Found. Phys.} {\bf 16}, 249 (1986).
\bibitem{CW}
P. Candelas and S. Weinberg,
{Nucl. Phys. B} {\bf 237}, 397 (1984).
\bibitem{Ra0}
M. Rainer,
{Multidimensional Scalar-Tensor Theories and Minisuperspace Approach}, 
publ.  in: Hadronic Journal {\bf 21}, 351 (1998); 
Proc.  Int.  Workshop 
on Modern Modified Theories of Gravitation and Cosmology, 
(Beer Sheva, June 29-30, 1997).
\bibitem{Ra1}
M. Rainer,
{Int. J. Mod. Phys. D} {\bf 4}, 397 (1995).
\bibitem{Ra2}
M. Rainer,
{Gravitation \& Cosmology} {\bf 1}, 121 (1995).
\bibitem{CEW}
E. J. Copeland, R. Easther and D. Wands,
{Phys. Rev. D} {\bf 56}, 874 (1997).
\bibitem{IMZ}
V. D. Ivashchuk, V. N. Melnikov and A. I. Zhuk,
{Nuovo Cimento B} {\bf 104}, 575 (1989).
\bibitem{Z}
A. Zhuk,
{Class. Quant. Grav.} {\bf 9}, 2029 (1992).
\bibitem{BIMZ}
U. Bleyer, V. D. Ivashchuk, V. N. Melnikov and A. Zhuk,
{Nucl. Phys. B} {\bf 429}, 177 (1994).
\bibitem{Z2}
A. Zhuk,
{Class. Quant. Grav.} {\bf 13}, 2163 (1996).
\bibitem{BZ1}
U. Bleyer and A. Zhuk,
{Gravitation \& Cosmology} {\bf 1}, 37 (1995);
ibid {\bf 1}, 106 (1995).
\bibitem{Ka}
E. Kasner,
{Am. J. Math.} {\bf 43}, 217 (1921).
\bibitem{ChD}
A. Chodos and S. Detweiler,
{Phys. Rev. D} {\bf 21}, 2167 (1980).
\bibitem{Fr}
P. G. O. Freund,
{Nucl. Phys. B} {\bf 209}, 146 (1982).
\bibitem{ACh}
T. Appelquist and A. Chodos,
{Phys. Rev. D} {\bf 28}, 772 (1983).
\bibitem{LP}
D. Lorenz-Petzold,
{Phys. Rev. D} {\bf 31}, 929 (1985).
\bibitem{KY}
T. Koikawa and M. Yoshimura,
{Phys. Lett. B} {\bf 155}, 137 (1985).
\bibitem{Ok}
Y. Okada,
{Nucl. Phys. B} {\bf 264}, 197 (1986).
\bibitem{Iv}
V. D. Ivashchuk,
{Phys. Lett. A} {\bf 170}, 16 (1992).
\bibitem{BlZ}
U. Bleyer and A. Zhuk,
{Astron. Nachr.} {\bf 317}, 161 (1996).
\bibitem{Le}
J. J. Levin,
{Phys. Lett. B} {\bf 343}, 69 (1995).
\bibitem{BG}
H. Bondi and T. Gold,
{Month. Not. R. Astron. Soc.} {\bf 108}, 252 (1948).


\end{thebibliography}
\end{document}